\documentclass[11pt]{article}

\usepackage{lineno,hyperref}
\modulolinenumbers[5]
\usepackage{bbm,amsmath,graphicx,natbib}
\usepackage{caption}
\usepackage{subcaption}
\usepackage{wrapfig}

\begin{document}
{\Large Distribution free testing of grouped Bernoulli trials}
\vspace{1em}
\begin{center}{\large Leigh A Roberts}\end{center}

\begin{center}{\it\small School of Economics and Finance, Victoria University of Wellington, \newline Wellington, New Zealand}\end{center}

\begin{center}Email: leigh.roberts@vuw.ac.nz\end{center}

\begin{center}10 January 2018\end{center}

\vspace{1em}
{\bf Abstract}
Recently Khmaladze has shown how to `rotate' one empirical process to another.
This paper is the first to apply 
this transform when successive data points are generated by a single distributional family, but with covariates varying over the sample.  
The application is to Bernoulli trials, and new results show how group sizes rotated are related to the number of parameters, and explore the impact of different types of data generating processes.
The utility of the rotation is clear: goodness of fit tests after rotation to a distribution free process are easily computed, show excellent convergence properties, and exhibit high power to reject incorrect null hypotheses.

\vspace{2em}
\section{Introduction}

In seeking to transform statistical goodness of fit tests into a distribution free format, consider first a centred empirical process having the basic form:
\begin{equation}
\frac1{\sqrt n}\sum_{i=1}^n\left[{\mathbbm{1}}_{\{\xi_i\leq y\}}-F_\theta(y)\right]
\label{e 11}
\end{equation}
in which $\mathbbm1$ is the indicator function; $\{\xi_i\}_{i=1}^n$ is the random sample of independent and identically distributed drawings; and $F_\theta(.)$ the distribution function in question. 
For fixed parameter $\theta$, the process in \eqref{e 11} is asymptotically a Brownian bridge in time $F_\theta$.  Standard goodness of fit tests such as the Kolmogorov Smirnov (KS) statistic may then be applied independently of the nature of the assumed distribution $F_\theta$, giving rise to so-called omnibus distribution free tests of the hypothesised or null distribution.

The extension of this testing procedure
to the case in which the parameter $\theta$ needs to be estimated from the data has been resolved by Khmaladze in two distinct ways.  As long ago as 1981, 
he proposed an alternative centring to the empirical process to produce a process with independent increments,
suitable for testing goodness of fit through a limiting process of a Brownian motion rather than a Brownian bridge: see \citet{khmaladze1981a}; and \citet{koul-swordson2011a}, together with references therein.

More recently,
\citet{khmaladze2016a} has shown how to `rotate' a given empirical process into a more or less arbitrary alternative empirical process; see also \citet{khmaladze2013a} where the theory is applied to discrete variables.
Evaluation of the utility of the new methodology is however in its infancy, with
\citet{nguyen-ttm2017a} apparently the only application of Khmaladze's rotations in the literature; see also \citet{nguyen-ttm2017b}.
The current paper is the first to apply this new transform when outcomes are generated by a single family of distributions, but depend on a known covariate varying over the sample \citep{khmaladze2017a}.

In the simple but practically important case of Bernoulli trials, we investigate the power to detect incorrect modelling of data, by calculating a KS statistic for the empirical process after rotation to a distribution free process.

Our basic approach is to generate four distinct processes of Bernoulli trials; we match these to four hypothesised models and fit these latter by maximum likelihood; we then transform each of these empirical processes to a common rotated process, for each of which we calculate a KS statistic; replicating the process allows us to calculate empirical distribution functions (EDFs) of the resulting KS statistics from the four transformed empirical processes.

The first experiment is to fit the correct model to each of the four underlying data generating processes.
The resulting EDFs are judged partly visually, and partly by calculation of empirical p-values after randomisation of pairs of EDFs.

The second experiment is to fit a common `logistic' model to each of the underlying processes, so that three of the hypothesised models are incorrect.  Judgement of the power to detect an incorrect null hypothesis is again partly by eye from the graphs of the EDFs, and partly by calculation of empirical p-values.

New results in this paper include a theorem 
facilitating the implementation of Khmaladze's rotation, which also relates the minimum group size to the number of parameters.  A further new result gives some insight into the bifurcation of results for different types of models.

In \S2 we set out the form of the empirical process underlying our work in this paper,
and apply Khmaladze's rotation of a stochastic process
to subgroups of sizes 1, 2 and 3.
We devote \S3 to setting up a matrix of score functions, and proving the theorems.

Simulation results are presented in \S 4.  
Restricting ourselves to a single covariate and a single parameter, four models are presented.  We discuss the way in which we obtain empirical p-values through randomisation.
Simulation results are discussed, 
for the first experiment concerning the validity of our transforms, and 
for the second experiment to determine the power to pick up incorrect null hypotheses.
A short conclusion summarises.

\section{Preamble}

\subsection{The empirical process}

Consider a sample of $nm$ independent Bernoulli trials, subdivided into $n$ groups each containing $m$ trials.
The $i$th trial within the $j$th subgroup is 
associated with a covariate $X_{ij}$, and produces a result
denoted $Y_{ij}$, 
for $1\leq i\leq m$, and $1\leq j\leq n$.
The $nm$ covariates $X_{ij}$ are mutually independent; and each variable $Y_{ij}$ assumes the value of 0 or 1, the former labelled a `failure' and the latter a `success'.
Set $X_j=(X_{1j},X_{2j},\ldots,X_{mj})$; and let $Z_j=(Y_{1j},Y_{2j},\ldots,Y_{mj})$ be the result from the $j$th subgroup, with realisations $z_j$.  It is convenient to order the elementary events from the $j$th subgroup `lexicographically': when $m=3$, for instance, $z_j=1,2,\ldots,8$ correspond to $(y_{1j},y_{2j},y_{3j})=000,001,010,\ldots,111$ respectively.  We have $1\leq z\leq 2^m$; and identify $z$ with $y$ when $m=1$.

We do not model the behaviour of the covariate $X$, and work with a joint distribution of the form
\begin{equation}
  P_\theta(dx,dz)=
  P_{x,\theta}(dz)H(dx)
\label{e 109}
\end{equation}
in which $\theta$ is a vector of $K$ parameters, and
the distribution $H$ is unspecified.
For the probability measure $P_{x,\theta}$, then, set $p_{x_{j},\,\theta}(z_{j})=P(Z_{j}=z_{j}\big|X_{j},\theta)$.
Denoting Borel sets in the respective spaces by $A$ and $B$,
we define the following centred empirical process:
\begin{equation}
\alpha^P_n(A,B,\theta)=
\frac1{\sqrt n}\sum_{j=1}^n\left[\mathbbm 1_{\{Z_j\in B\}}-P_\theta(B\big|X_j)\right]\mathbbm 1_{\{X_j\in A\}}
\label{e 108}
\end{equation}
It is convenient to
parametrise processes by functions. 
Considering functions $\phi(x,z)$ in $L_2(P_\theta)$, the space of square integrable functions with respect to the measure $P_\theta$, 
we rewrite \eqref{e 108} in the form
\begin{equation}
\alpha_n^P(\phi,\theta)=
\int\int\phi(x,z)\alpha^P_n(dx,dz,\theta)
\label{e 114}
\end{equation}

We wish to transform processes defined on $L_2(P_\theta)$ to processes defined on say $L_2(Q_\theta)$, where by analogy to \eqref{e 109} we have
\[
Q_\theta(dx,dz)=Q_{x,\theta}(dz)H(dx)
\]
Consider functions $\phi\in L_2(P_\theta)$ and $\psi\in L_2(Q_\theta)$.
Provided that $P_\theta$ and $Q_\theta$ are equivalent, define
\begin{equation}
\ell_{x,\theta}(dz)=\sqrt{\frac{dQ_{x,\theta}(dz)}{dP_{x,\theta}(dz)}}
\label{e 110}
\end{equation}
so that $\psi\in L_2(Q_\theta)$ if and only if $\ell\psi\in L_2(P_\theta)$.  
Since the functions $\phi$ and $\psi$ are square integrable in $(x,z)$ with respect to $P_{\theta}$ and $Q_{\theta}$ respectively, they are square integrable in $z$ with respect to
$P_{x,\theta}$ and $Q_{x,\theta}$ respectively, for $H-$almost all $x$.
The precise form of $H$ being unspecified, it is convenient to assume that the functions $\phi$ and $\psi$ are bounded in $x$; since $z$ assumes but finitely many values, the functions are bounded in both $x$ and $z$.

Our intention is to simplify the conditional $Q_{x,\theta}(z)$ measure so that the numerator of the quantity under the square root in \eqref{e 110} depends on neither $x$ nor $\theta$: \eqref{e 110} becomes
\begin{equation}
\ell_{x,\theta}(z)=\sqrt{\frac{\prod_{i=1}^{m}q^{1-y_i}(1-q)^{y_i}}{\prod_{i=1}^{m}p_{x_i,\theta}^{1-y_i}(1-p_{x_i,\theta})^{y_i}}}
\label{e 119}
\end{equation}
with $q$ and $p_{x_i,\theta}$ denoting probabilities of failure, and $y_i=1$ indicating success.

\subsection{The Kolmogorov Smirnov goodness of fit test}

The transformation of the empirical process on $L_2(P_{x,\theta})$ to that on $L_2(Q)$ is effected as follows.
From  \citet{khmaladze2016a} and \eqref{e 114},
\begin{equation}
\alpha_n^Q(\psi)=\alpha_n^P(U(\ell\psi),\theta)
\label{e 111}
\end{equation}
in which $U$ is a unitary operator, to be specified below.

We wish to carry out goodness of fit tests as to whether or not the data in \eqref{e 108} could reasonably have been generated by $P_\theta$, for some value of $\theta$, assumed for the moment to be known.
A standard means of doing this would be to employ the KS statistic, defined from \eqref{e 114} as
\[
\underset{\phi\in\Phi}\max\ |\alpha_n^P(\phi,\theta)|
\]
where 
the maximum is taken over all functions $\phi$ in a class $\Phi$ of functions in $L_2(P_\theta)$.
The point of transforming to the $Q$ space, however, is that we may instead employ the KS statistic in the $Q$ space:
\[
\underset{\psi\in\Psi}\max\ |\alpha_n^Q(\psi)|
\]
and test goodness of fit by an empirical process defined on a simpler measure $Q$ depending on neither $x$ nor $\theta$.

A suitable class $\Psi$ is generated by indicator functions in $x$ and $z$:
\[
\psi_{x_0,z_0}(x,z)=\mathbbm1_{\{x\leq x_0\}}\left[\mathbbm1_{\{z\leq z_0\}}-2^{-m}\,z_0\right]
\]
in which $\psi$ is a vector of length $2^m$, so that the operator $U$ in \eqref{e 111} assumes the form of a $2^m\times2^m$ orthogonal matrix.

\subsection{The nature of the operator $U$}

Setting $a_0=b_0=\mathbf1$ to be vectors consisting entirely of ones, the first requirement for the operator $U$ in \eqref{e 111} is that $U(\ell\,b_0)=a_0$, with $\ell$ given in \eqref{e 119}.
One further constraint arises for each unknown parameter $\theta_k$, to be estimated by maximum likelihood.
Choosing vectors $b_k$, for $1\leq k\leq K$, such that $\{b_k\}_0^K$ are mutually orthonormal with respect to the probability measure $Q$, $U$ must simultaneously map $\ell\,b_k$ to $a_k$, for $k=0,\ldots,K$.  

The construction of a unitary operator $U$ satisfying these properties can be accomplished as a sequence of reflections in two dimensional subspaces:  see \citet{khmaladze2013a} and \citet[p.\ 67]{nguyen-ttm2017b}.  Instead we apply Theorem 2 below, which seems a more direct approach, especially for large numbers of parameters.

\section{The score function matrix and theorems}

The (non-normalised) score function for the $k$th parameter and the $j$th subgroup is $\left(\partial\big/\partial\theta_{k}\right) p^{(j)}_z\big/p^{(j)}_z$, in which $p^{(j)}_z=p_{x_{j},\,\theta}(z_{j})$.

We bind these score functions together as a $2^m\times K$ matrix, say $M^{(j)}$, with the $\left(\partial\big/\partial\theta_{k}\right)$ terms appearing in the $k$th column and the rows indexed by $z$.
Further defining the $2^m\times 2^m$ diagonal matrix $D^{(j)}_P$ to contain $p^{(j)}_z$ in order down the diagonal, the $K\times K$ information matrix is given as
\begin{equation}
\Gamma^{(j)}={M^{(j)}}^T\,D^{(j)}_P\,M^{(j)}
\label{e 82}
\end{equation}
The information matrix for the whole sample is $\sum_{j=1}^n\Gamma^{(j)}$.  Finally, from \eqref{e 119} we define $L^{(j)}$ as a $2^m\times 2^m$ diagonal matrix containing $\ell_{x_{j},\theta}(z)$.  From now on we omit the index $j$.

The normalised score functions are the columns $a_k$ of $M\,\Gamma^{-1/2}$.  Define the $2^m\times (K+1)$ matrix $A=(\mathbf1|M\,\Gamma^{-1/2})$, with columns $\{a_k\}_{k=0}^K$.  We further define $O_A={D_P}^{1/2}\,A$, and let $I_r$ denote the identity matrix of order $r$.

{\bf Lemma 1}

Using the above definitions, and provided $K+1\leq2^m$, we have
\[
O_A^T\,O_A=I_{K+1}
\]

{\bf Proof}

The information matrix $\Gamma$ is symmetric and positive definite, so that its square root and the inverse can be defined via the spectral decomposition.  Then considering $O_A^T\,O_A$ as a $2\times2$ block matrix, 
differentiating $\sum p_z=1$ yields that $\mathbf1^T\,D_P\,\,M=0$; and from \eqref{e 82}
\[
\ \hspace{10em}\Gamma^{-1/2}\,M^T\,D_P\,M\,\Gamma^{-1/2}=I_K\hspace{10em}\diamond
\]

When $K+1<2^m$ we insert into $O_A$ additional columns, chosen arbitrarily from the orthogonal complement to the column space of $O_A$, to produce an orthogonal matrix $O_P$, say $O_P=D_P^{1/2}\,(A|Z_A)$.
When $K+1=2^m$ we simply set $O_P=O_A$.

We analogously define the orthogonal matrix $O_Q=D_Q^{1/2}\,(B|Z_B)$, in which $D_Q=2^{-m}\,I_{2^m}$.  While we regard $B$ as a $2^m\times(K+1)$ matrix, and may refer to its columns as $\{b_k\}_{0}^K$, only the first column of $B$ is specified as $\mathbf1$: the remaining columns of $O_Q$ may be chosen arbitrarily, subject to the orthogonality constraints.

{\bf Theorem 2}

In the above notation:\\
a.  When $K+1>2^m$, not all of the $K$ unknown parameters can be identified, and the rotation of subgroups of $m$ sample values is not defined.\\
b.  When $K+1\leq 2^m$, the unitary transform $U$ in \eqref{e 111} may be taken as
\[
U=D_P^{-1/2}\,O_P\,O_Q^{-1}\,D_P^{1/2}
\]

{\bf Proof}

Noting that $L=D_Q^{1/2}\,D_P^{-1/2}$,
\[
UL(B|Z_B)=D_P^{-1/2}\,O_P\,O_Q^{-1}\,D_P^{1/2}L(B|Z_B)=D_P^{-1/2}\,O_P\,O_Q^{-1}\,D_Q^{1/2}(B|Z_B)=(A|Z_A)
\]
so that in particular $ULB=A$.\hfill$\diamond$

For applications of Theorem 1 in this paper, we assume $b_1^T$ to be equal to
\[
(-1,1)\qquad\sqrt2(-1,0,0,1)\qquad\frac1{\sqrt{3}}(-3,-1,-1,1,-1,1,1,3)
\]
for $m=1,2$ and 3 respectively.

{\bf Lemma 3}

Assuming a single covariate $x$ and a single parameter $\theta$, let $p_k=p_{x,\theta}(k)$
and $\dot p_k=(\partial/\partial\theta)p_k\big|_{\theta=\hat\theta}$, for $k=0,1$; 
and finally set $m=1$.  

Then the score function $a_1$ reduces to $(\pm\sqrt{p_1/p_0},\mp\sqrt{p_0/p_1})^T$ and the transform $U$ in \eqref{e 111} is given 
below, the upper sign corresponding to $\dot p_0>0$, and the lower sign to $\dot p_0<0$:
\[
U=\frac1{\sqrt2}\begin{pmatrix}\sqrt{p_0}\left(1\mp\sqrt{p_1/p_0}\right)&\sqrt{p_1}\left(1\pm\sqrt{p_1/p_0}\right)\\\sqrt{p_0}\left(1\pm\sqrt{p_0/p_1}\right)&\sqrt{p_1}\left(1\mp\sqrt{p_0/p_1}\right)\end{pmatrix}
\]

{\bf Proof}

It may be verified that $U(\ell b_k)=a_k$, for $k=0,1$.
\hfill$\diamond$

{\bf Theorem 4}

Retaining the assumptions and notation from Lemma 3, define $p_k^{(r)}$ and $\dot p_k^{(r)}$ for $k=0,1$, as well as score functions $a_1^{(r)}$, for models $r=m1,m2$.  Suppose either that $Pr\left(a_1^{(m1)}=a_1^{(m2)}\right)=1$ or that $Pr\left(a_1^{(m1)}=-a_1^{(m2)}\right)=1$, where the probability is defined over the random variables $x$ and $\hat\theta$.  

Then the empirical processes for models $m1$ and $m2$ rotate to the same distribution in the $Q$ space.

{\bf Proof}

Suppose the first condition applies.  Then the proof of \eqref{e 111} in \citet[Thm.\ 7]{khmaladze2016a} applies because $U$ is well defined.  In the second case, simply reverse the sign of one of the parameters to obtain the first situation.\hfill$\diamond$

\begin{figure}
\centering
\begin{subfigure}{.5\textwidth}
  \centering
  \includegraphics[width=1.1\linewidth]{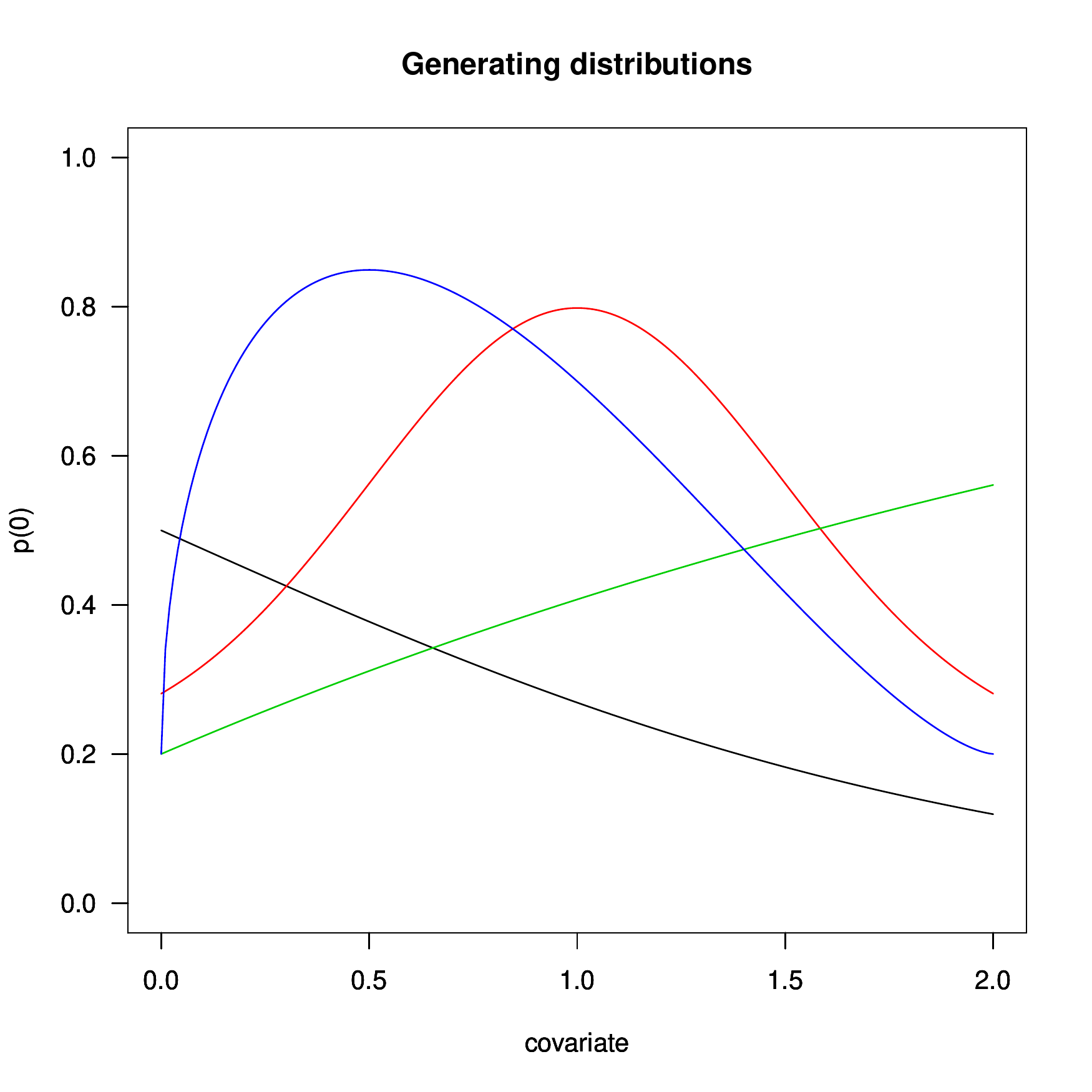}
\end{subfigure}%
\begin{subfigure}{.5\textwidth}
  \centering
  \includegraphics[width=1.1\linewidth]{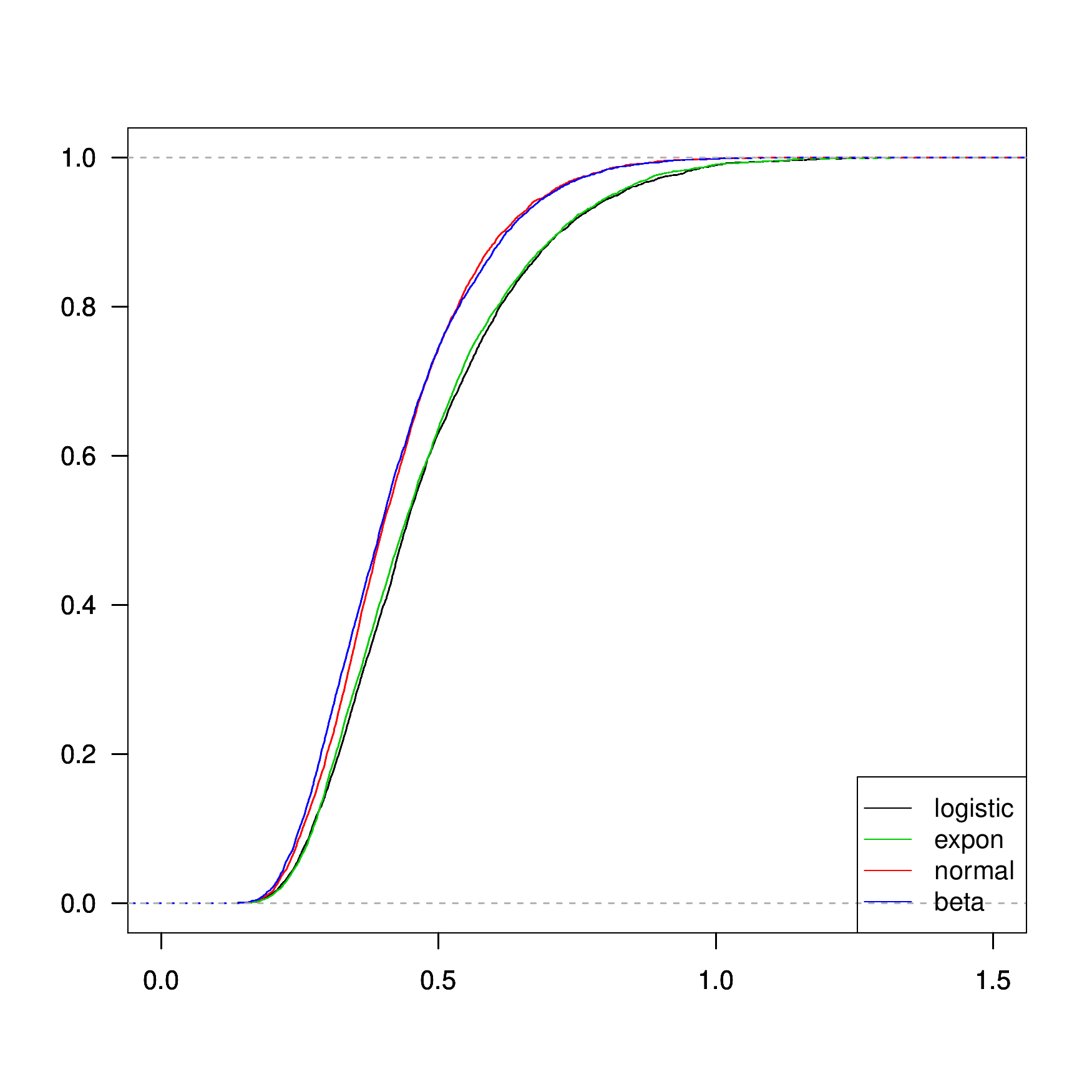}
\end{subfigure}
\caption{Failure probabilities for the four models, and results of the first experiment for $m=1$}
\label{Rg299-313b}
\end{figure}

\begin{figure}
\centering
\begin{subfigure}{.5\textwidth}
  \centering
  \includegraphics[width=1.1\linewidth]{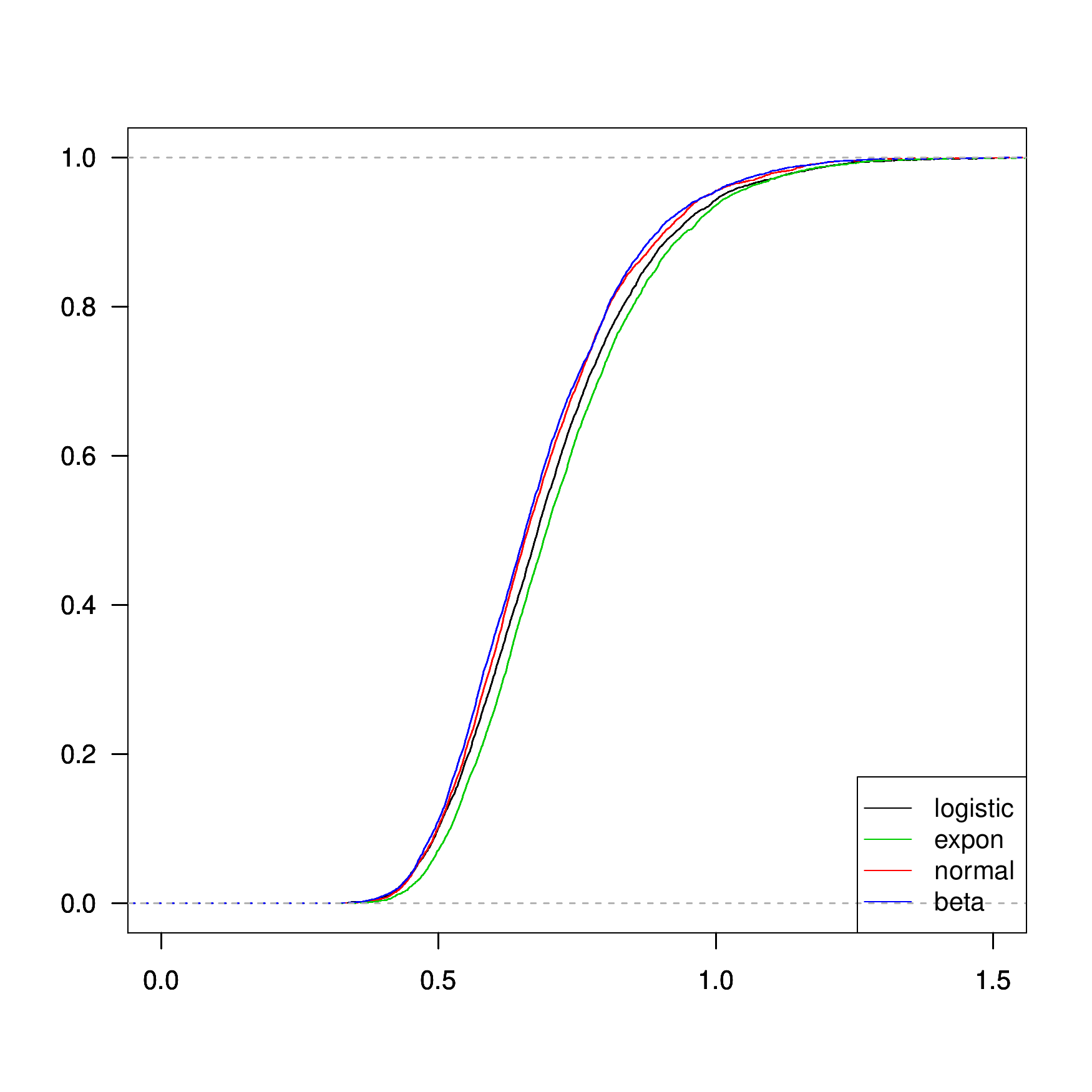}
\end{subfigure}%
\begin{subfigure}{.5\textwidth}
  \centering
  \includegraphics[width=1.1\linewidth]{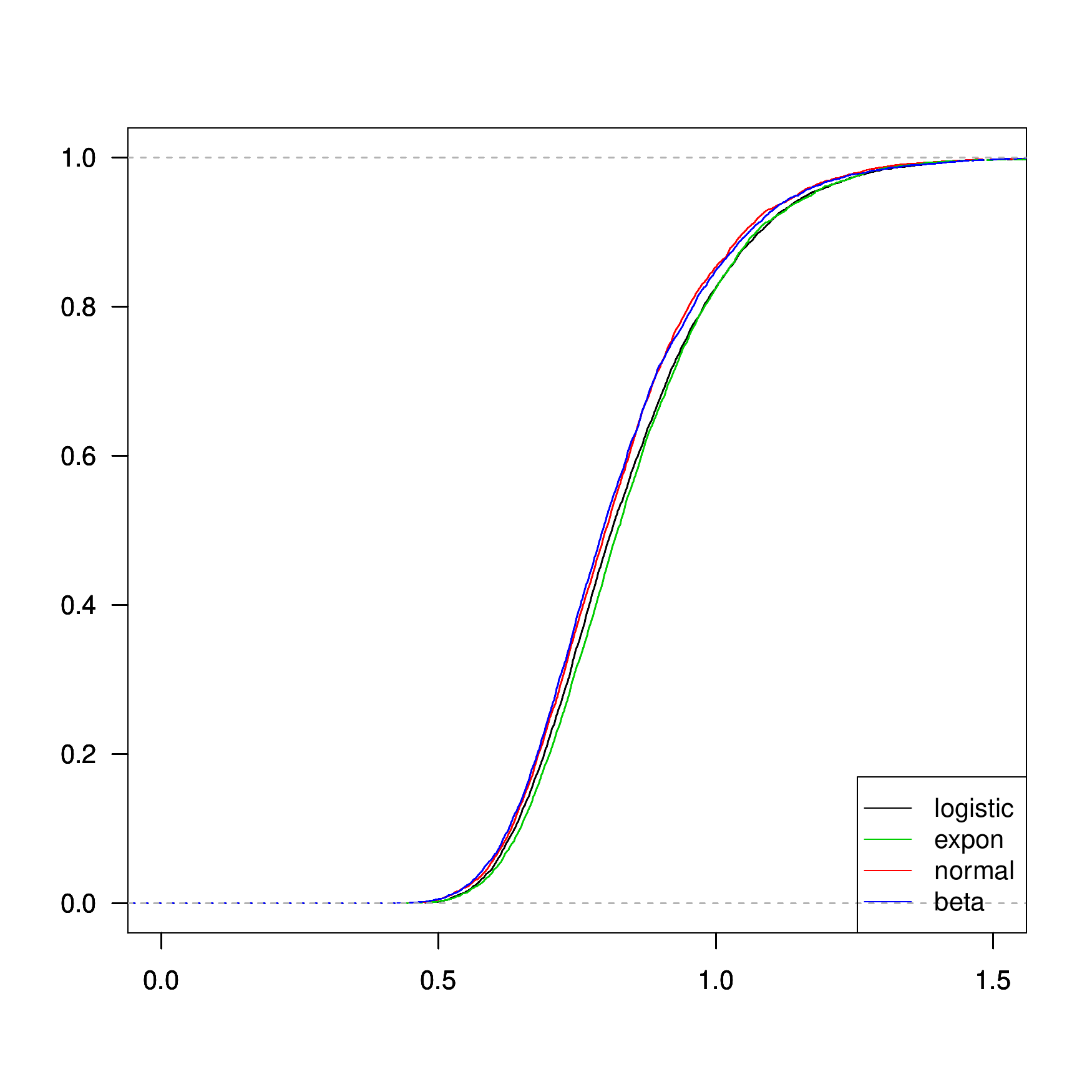}
\end{subfigure}
\caption{Results of the first experiment for $m=2$ (on the left) and $m=3$}
\label{Rg310b-309b}
\end{figure}

\section{Simulation of Bernoulli trials}

The four $P_\theta$ measures in \S\ref{s 1} are chosen to generate the data, using the given parameters $\theta_0$.  The same measures are then used to test goodness of fit, estimating unknown parameters $\theta$ by maximum likelihood.
The next section discusses what is meant in this paper by empirical p-values, following which we present the results of the simulations in the first and second experiments.

Covariates $X_{ij}$ are chosen uniformly between 0 and 2, for $1\leq i\leq m,\ 1\leq j\leq n$.  
In calculating KS statistics, we have used a grid size of 100 intervals between 0 and 2 for the covariates.
Sample size throughout is 96 Bernoulli trials; and the number of replications for each EDF is 5000.

\subsection{The four models\label{s 1}}

Figure \ref{Rg299-313b}(a) indicates several functions for mapping the covariate $x$ to the probability of failure $p_{x,\theta_0}(0)$.  The downward sloping solid line shows the logistic distribution, with failure probability:
\begin{equation}
p_{x,\theta}(0)=\frac1{1+e^{\theta\,x}}\qquad\mbox{with}\ \theta_0=1
\label{e 94b}
\end{equation}
which provides the first of our data generating models, and the default fitted model for the investigation of power in \S\ref{s 3}.
For the second data generating process
we rescale the distribution function of the exponential distribution:
\begin{equation}
p_{x,\theta}(0)=0.2+0.8(1-e^{-\theta\,x})\qquad\mbox{with}\ \theta_0=0.3
\label{e 103}
\end{equation}
which provides the upward sloping graph in Figure \ref{Rg299-313b}(a). 
The third data generating model is furnished by
a modified normal density:
\begin{equation}
  p_{x,\theta}(0)=\frac{3/2}{\sqrt{2\,\pi}}\,\exp\left[-2(x-\theta)^2\right]
  +0.2\qquad\mbox{with}\ \theta_0=1
\label{e 100}
\end{equation}
Finally, our fourth data generating process is based on the positively skewed beta density
\begin{equation}
p_{x,\theta}(0)=0.2+2\left(\frac x2\right)^{0.5}\left(1-\frac x2\right)^{\theta-1}\qquad\mbox{with}\ \theta_0=2.5
\label{e 101}
\end{equation}

\subsection{Empirical p-values}

We wish to test whether a given pair of EDFs, say $G_1$ and $G_2$, arise from the same underlying distribution.
Given some measure of distance between $G_1$ and $G_2$, say $d(G_1,G_2)$, we draw randomly from the combined sample values to obtain two new EDFs, and calculate the distance between these artificial EDFs; and we do this $N$ times say.  Then we calculate the position of $d(G_1,G_2)$ in the ordered sample of $N$ simulated distances, to find an empirical p-value \citep[e.g.][p.\ 4]{manly2007a}.  We test for commonality of the EDFs of the KS statistics arising from the four transformed empirical processes, with distance between distribution functions taken as the maximum vertical distance between them.  For the results in Table 1, we used $N=10,000$.

\subsection{First experiment}

The agreement of EDFs predicted by theory is largely confirmed for the second pair of models in \S\ref{s 1} (the `normal' and `beta' models) to judge from Figures \ref{Rg299-313b}(b) and \ref{Rg310b-309b}.  
For grouped data, the matching of the EDFs in Figure \ref{Rg310b-309b} is borne out by the empirical p-values in Table 1. 
When $m=1$ however, for each of the third and fourth models presented in \S\ref{s 1}, from Lemma 3 a given value of $p_0$ may arise from either $\dot p_0>0$ or $\dot p_0<0$, depending on the values of $x$ and $\hat\theta$ generating the value of $p_0$.
The ambiguity in the transform $U$ 
is reflected in Table 1, although the visual agreement between the corresponding EDFs in Figure \ref{Rg299-313b}(b) `seems' reasonable.

For the first pair of models in \S\ref{s 1}, the situation is reversed.  For $m=1$, Theorem 4 applies (the second condition is satisfied), and Table 1 confirms the matching of the corresponding EDFs.  The bifurcation in Figure\ref{Rg299-313b}(b) between the two types of models is clear.

When $m>1$, 
for the second pair of models in \S\ref{s 1} the only constraint on the matrix $A$ in Lemma 1, apart from the orthonormality of the columns, is that the first column is a vector of ones; whereas further constraints are imposed for each of the first two models.  The result is that for $m>1$, the EDFs for the second pair of models agree well according to Table 1, but the EDFs for the first pair diverge.

Notwithstanding empirical p-values, however, 
visual inspection of Figures \ref{Rg299-313b}(b) and \ref{Rg310b-309b} seems to show close agreement between the EDFs of the first two models.

\begin{table}\begin{center}
\begin{tabular}{|c|c|ccc|}\hline
First experiment&m&Model 2&Model 3&Model 4\\\hline
Model 1&1&0.153&0.000&0.000\\
       &2&0.000&0.000&0.000\\
       &3&0.005&0.000&0.000\\\hline
Model 2&1&   &0.000&0.000\\
       &2&   &0.000&0.000\\
       &3&   &0.000&0.000\\\hline
Model 3&1&   &   &0.001\\
       &2&   &   &0.074\\
       &3&   &   &0.509\\\hline
\end{tabular}
\caption{Empirical p-values for the first experiment}
\label{t 1}
\end{center}\end{table}

\subsection{Second experiment\label{s 3}}
\begin{figure}
\includegraphics[height =1.7in, width=4in,angle=0]{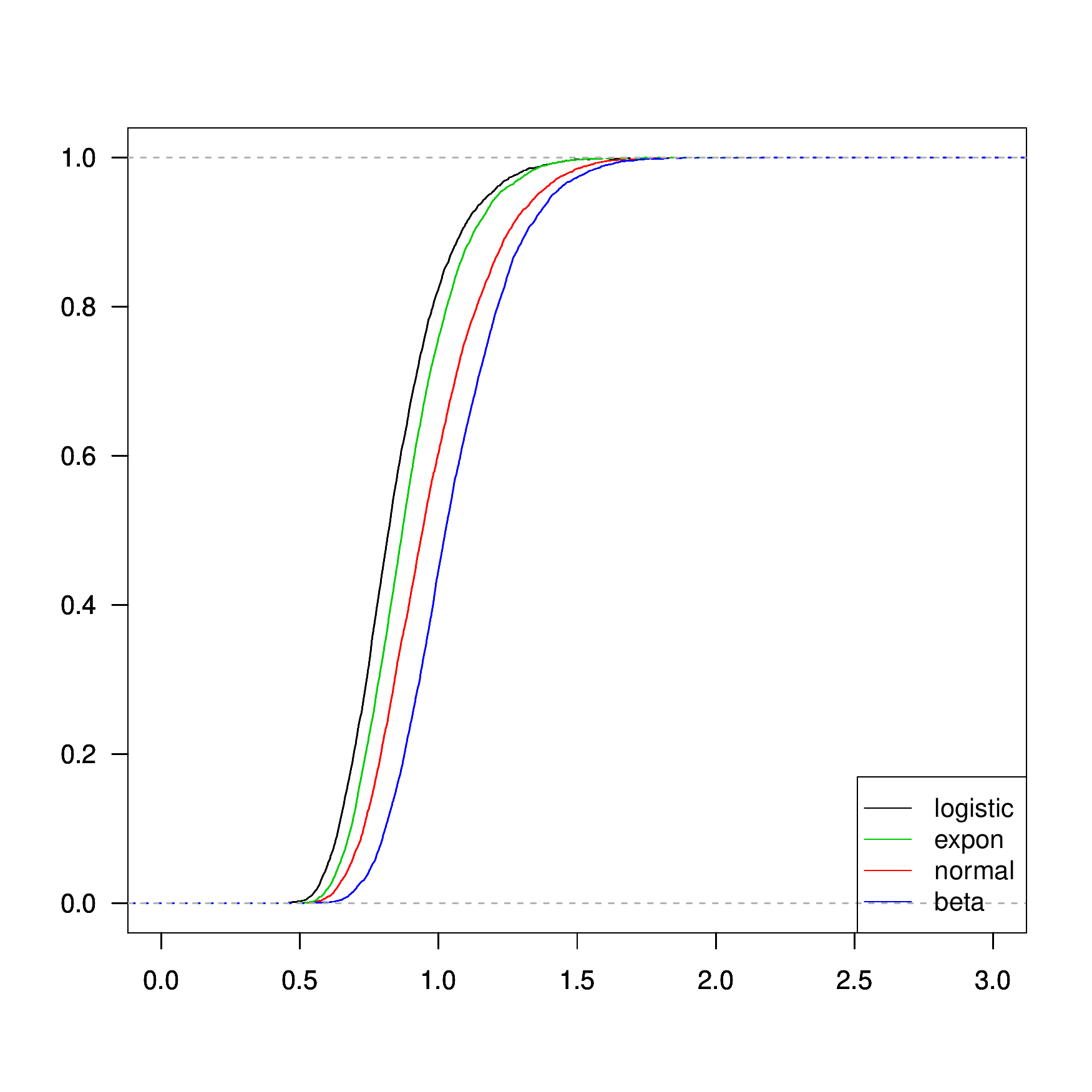}
\caption{Results of the second experiment: power when $m=3$}
\label{Rg308a}
\end{figure}

Regardless of the different natures of the two pairs of models, the power to detect when an incorrect model is fitted seems excellent.  The EDFs are well separated for group sizes of 1 and 2; and even though the visual difference is much reduced when $m=3$, the empirical p-values between all pairs of EDFs for the four models in Figure \ref{Rg308a} are all zero.

\begin{figure}
\includegraphics[height =1.7in, width=4in,angle=0]{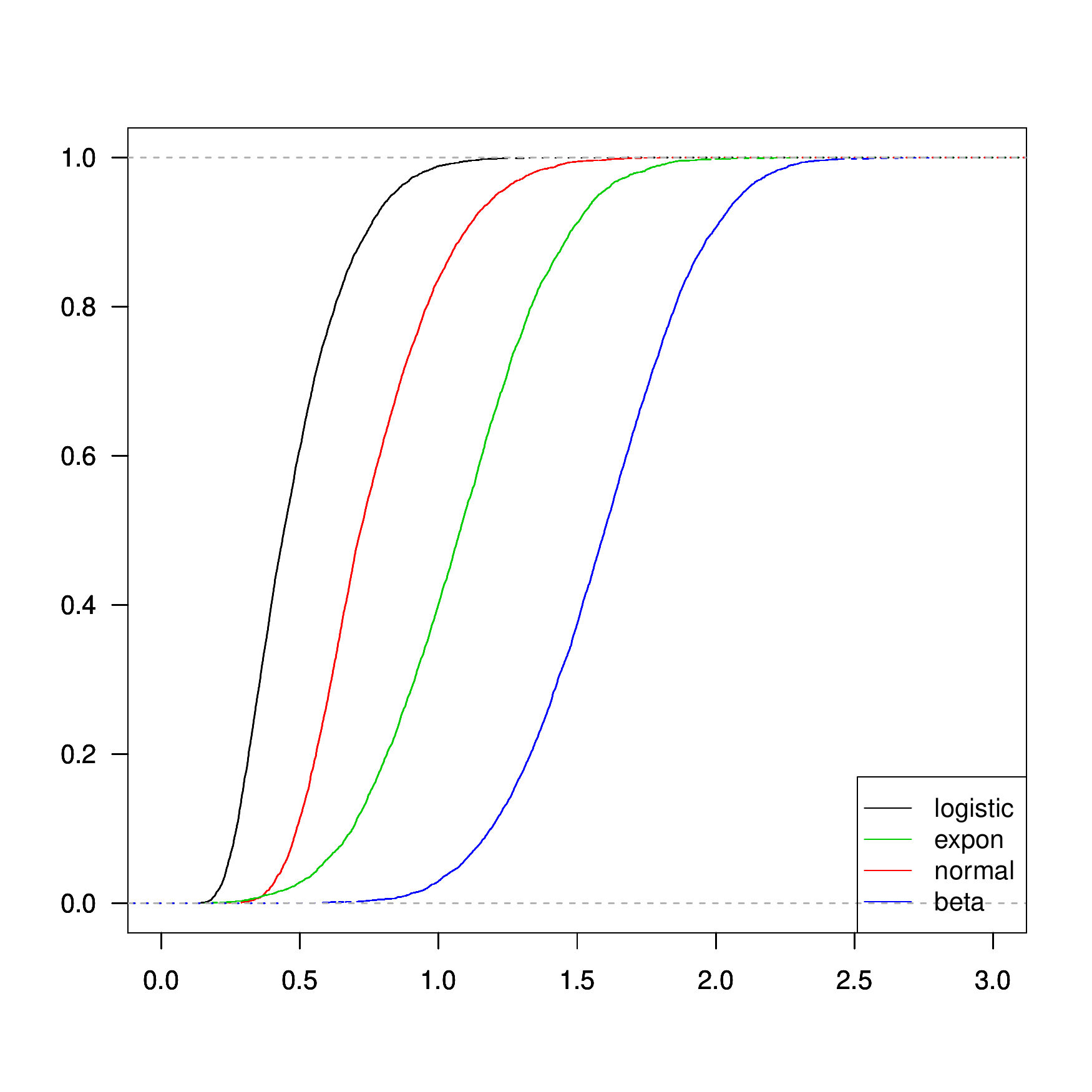}
\caption{Results of the second experiment: power when $m=1$}
\label{Rg314a}
\end{figure}

\begin{figure}
\includegraphics[height =1.7in, width=4in,angle=0]{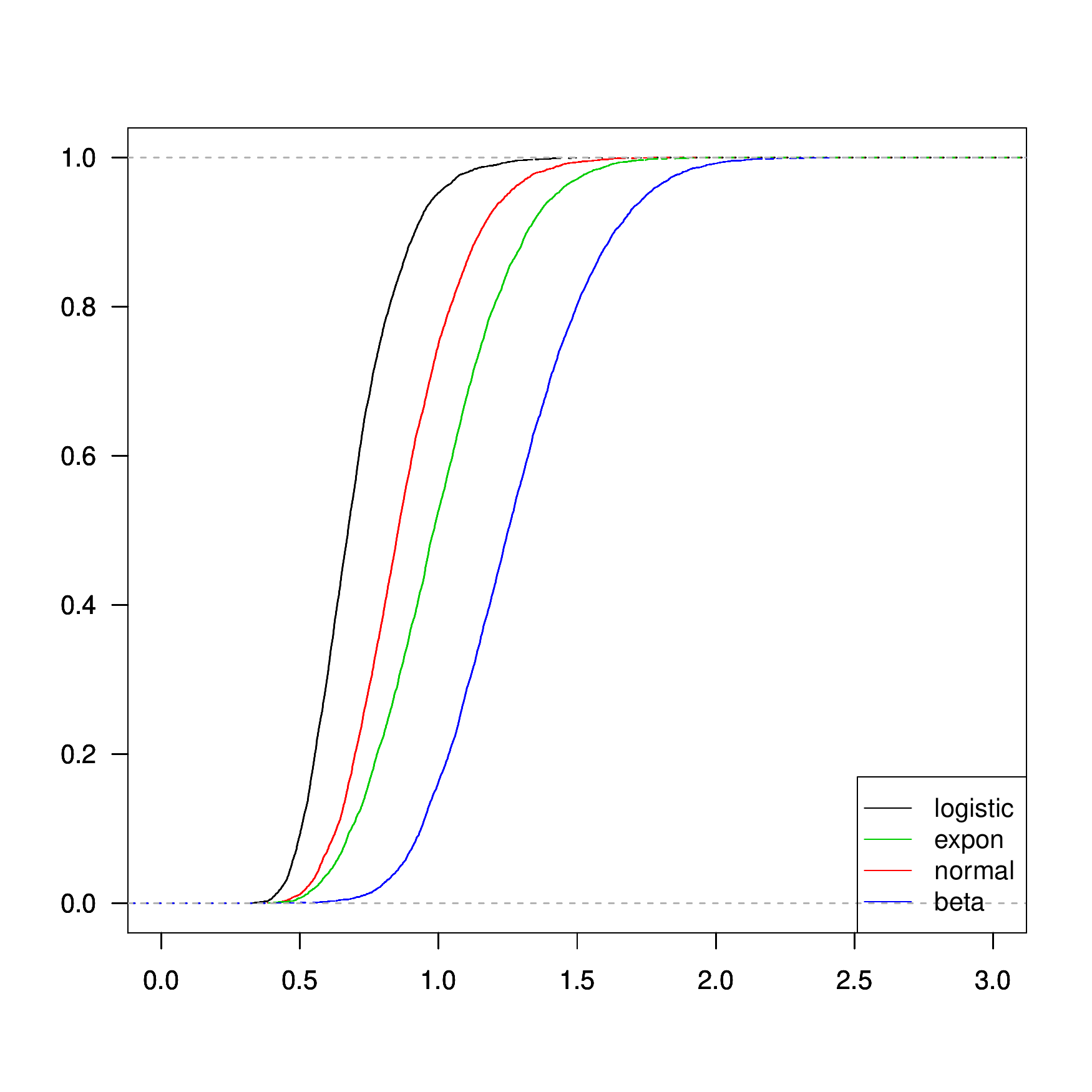}
\caption{Results of the second experiment: power when $m=2$}
\label{Rg311a}
\end{figure}

\section{Conclusion}

For distribution free goodness of fit tests in the context of Bernoulli trials, 
we have applied Khmaladze's method of rotating empirical processes to a simpler process.

In addition to establishing a relationship between size of group rotated and the number of parameters to be estimated, we have also facilitated the implementation of the rotation, and shown that the properties of the distribution free tests depends on the nature of constraints on the score function of the underlying data generating processes.

Much work clearly needs to be done to investigate the extent to which it is desirable to work with groups of different sizes, and to investigate further the way in which the nature of the underlying models impinges on tests of goodness of fit, for Bernoulli trials and for more general classes of models.
Nevertheless, the utility of the transform of empirical processes as set out in \citet{khmaladze2016a} has clearly been shown.  Moreover, implementation of the procedure is straightforward, in addition to which the computational burden is modest, and the power to detect incorrectly hypothesised models is high.

\section*{Acknowledgement}Thanks to Estate Khmaladze for comments on early drafts of this paper.  Responsibility for errors naturally remains with the author.

\bibliography{}

\begin{thebibliography}{8}
\expandafter\ifx\csname natexlab\endcsname\relax\def\natexlab#1{#1}\fi
\providecommand{\url}[1]{\texttt{#1}}
\providecommand{\href}[2]{#2}
\providecommand{\path}[1]{#1}
\providecommand{\DOIprefix}{doi:}
\providecommand{\ArXivprefix}{arXiv:}
\providecommand{\URLprefix}{URL: }
\providecommand{\Pubmedprefix}{pmid:}
\providecommand{\doi}[1]{\href{http://dx.doi.org/#1}{\path{#1}}}
\providecommand{\Pubmed}[1]{\href{pmid:#1}{\path{#1}}}
\providecommand{\bibinfo}[2]{#2}
\ifx\xfnm\relax \def\xfnm[#1]{\unskip,\space#1}\fi
\bibitem[{Khmaladze(1981)}]{khmaladze1981a}
\bibinfo{author}{Khmaladze, E.V.}, \bibinfo{year}{1981}.
\newblock \bibinfo{title}{Martingale approach in the theory of goodness-of-fit
  tests}.
\newblock \bibinfo{journal}{Theory of Probability and its Applications}
  \bibinfo{volume}{26}, \bibinfo{pages}{240--257}.
\bibitem[{Khmaladze(2013)}]{khmaladze2013a}
\bibinfo{author}{Khmaladze, E.V.}, \bibinfo{year}{2013}.
\newblock \bibinfo{title}{Note on distribution free testing for discrete
  distributions}.
\newblock \bibinfo{journal}{Annals of Statistics} \bibinfo{volume}{41},
  \bibinfo{pages}{2979--2993}.
\bibitem[{Khmaladze(2016)}]{khmaladze2016a}
\bibinfo{author}{Khmaladze, E.V.}, \bibinfo{year}{2016}.
\newblock \bibinfo{title}{Unitary transformations, empirical processes and
  distribution free testing}.
\newblock \bibinfo{journal}{Bernoulli} \bibinfo{volume}{22},
  \bibinfo{pages}{563--588}.
\bibitem[{Khmaladze(2017)}]{khmaladze2017a}
\bibinfo{author}{Khmaladze, E.V.}, \bibinfo{year}{2017}.
\newblock \bibinfo{title}{Distribution free testing for conditional
  distributions given covariates}.
\newblock \bibinfo{journal}{Statistics \& Probability Letters}
  \bibinfo{volume}{129}, \bibinfo{pages}{348--354}.
\bibitem[{Koul and Swordson(2011)}]{koul-swordson2011a}
\bibinfo{author}{Koul, H.L.}, \bibinfo{author}{Swordson, E.},
  \bibinfo{year}{2011}.
\newblock \bibinfo{title}{Khmaladze transformation}, in:
  \bibinfo{booktitle}{International Encyclopedia of Statistical Science}.
  \bibinfo{publisher}{Springer}, pp. \bibinfo{pages}{715--718}.
\bibitem[{Manly(2007)}]{manly2007a}
\bibinfo{author}{Manly, B.F.J.}, \bibinfo{year}{2007}.
\newblock \bibinfo{title}{Randomization, Bootstrap and Monte Carlo Methods in
  Biology}.
\newblock \bibinfo{publisher}{Chapman \& Hall}.
\bibitem[{Nguyen(2017a)}]{nguyen-ttm2017b}
\bibinfo{author}{Nguyen, T.T.M.}, \bibinfo{year}{2017}a.
\newblock \bibinfo{title}{Asymptotic methods of testing statistical
  hypotheses}.
\newblock Ph.D. thesis. School of Mathematics and Statistics, Victoria
  University, Wellington, New Zealand.
\bibitem[{Nguyen(2017b)}]{nguyen-ttm2017a}
\bibinfo{author}{Nguyen, T.T.M.}, \bibinfo{year}{2017}b.
\newblock \bibinfo{title}{A new approach to distribution free tests in
  contingency tables}.
\newblock \bibinfo{journal}{Metrika} \bibinfo{volume}{80},
  \bibinfo{pages}{153--170}.

\end{thebibliography}

\end{document}